# Black Phosphorus Field-effect Transistors

Likai Li[1], Yijun Yu[1], Guo Jun Ye[2], Qingqin Ge[1], Xuedong Ou[1], Hua Wu[1], Donglai Feng[1], Xian Hui Chen[2*] and Yuanbo Zhang[1*]

[1]*State Key Laboratory of Surface Physics and Department of Physics, Fudan University, Shanghai 200438, China*

[2]*Hefei National Laboratory for Physical Science at Microscale and Department of Physics, University of Science and Technology of China, Hefei, Anhui 230026, China*

*Email: chenxh@ustc.edu.cn, zhyb@fudan.edu.cn

**Two-dimensional crystals have emerged as a new class of materials with novel properties that may impact future technologies. Experimentally identifying and characterizing new functional two-dimensional materials in the vast material pool is a tremendous challenge, and at the same time potentially rewarding. In this work, we succeed in fabricating field-effect transistors based on few-layer black phosphorus crystals with thickness down to a few nanometers. Drain current modulation on the order of $10^5$ is achieved in samples thinner than 7.5 nm at room temperature, with well-developed current saturation in the I-V characteristics – both are important for reliable transistor performance of the device. Sample mobility is also found to be thickness dependent, with the highest value up to ~ 1000 $cm^2$/Vs obtained at thickness ~ 10 nm. Our results demonstrate the potential of black phosphorus thin crystal as a new two-dimensional material for future applications in nano-electronic devices.**

Black phosphorus is a layered material in which individual atomic layers are stacked together by Van der Waals interactions, much like bulk graphite[1]. Inside a single layer, each phosphorus atom is covalently bonded with three adjacent phosphorus atoms to form a puckered honeycomb structure[2–4] (Fig. 1a). The three bonds take up all three valence electrons of phosphorus, so unlike graphene[5,6] a monolayer black phosphorus (termed "phosphorene") is a semiconductor with a predicted direct band gap of ~ 2 eV at the $\Gamma$ point of the first Brillouin zone[7]. For few-layer phosphorene, interlayer interactions reduce the band gap for each layer added, and eventually reach ~ 0.3 eV (refs 8–11) for bulk black phosphorus. The direct gap also moves to the **Z** point as a consequence[7,12]. Such a band structure provides a much needed gap for the field-effect transistor (FET) application of two dimensional (2D) materials such as graphene, and the thickness-dependent direct band gap may lead to potential applications in optoelectronics, especially in the infrared regime. In addition, observations of phase transition from semiconductor to metal[13,14] and superconductor under high pressure[15,16] indicate correlated phenomena play an important role in black phosphorus under extreme conditions. Here we fabricate few-layer phosphorene devices and study their electronic properties modulated by electric field effect. Excellent transistor performances are achieved at room temperature. In particular, important metrics of our devices such as drain current modulation and mobility are either better or comparable to FETs based on other layered materials[17,18].

We start with bulk black phosphorus crystals that we grow under high pressure and at high temperatures (see Methods). The band structure of bulk black phosphorus is verified by angle-resolved photoemission spectroscopy (ARPES) measurements as well as *ab initio* calculations. The filled bands of freshly cleaved bulk crystal as measured by ARPES are shown in Fig. 1b, which for the most part agree with screened hybrid functional calculations with no material-dependent empirical parameters (dashed and solid lines in Fig. 1b for filled and empty bands respectively). The calculated band gap (~ 0.2 eV) also agrees reasonably well with previous measurements[8–11], considering screened hybrid functional calculations tend to underestimate slightly the size of the band gap in semiconductors[19–21].

We next fabricate few-layer phosphorene FETs with a back-gate electrode (see Fig. 2a). A scotch tape based mechanical exfoliation method is employed to peel thin flakes from bulk crystal onto degenerately doped silicon wafer covered with a layer of thermally grown silicon dioxide. Optical microscopy and atomic force microscopy (AFM) are used to hunt thin flake samples and determine their thickness (Fig. 2a). Metal contacts are then deposited on black phosphorus thin flakes by sequential electron-beam evaporation of Cr and Au (typically 5 nm and 60 nm, respectively) through a stencil mask that is aligned with the sample. Standard electron beam lithography process and other contact metals such as Ti/Au are also used to fabricate few-layer phosphorene FETs, and similar results are obtained in terms of device performance.

The switching behaviour of our few-layer phosphorene transistor at room temperature is characterized in vacuum (~ $10^{-6}$ mBar), in a configuration depicted in Fig. 2a. We sweep the back-gate voltage, $V_g$, applied to the degenerated doped silicon, as the source-drain bias, $V_{ds}$, across the black phosphorus conductive channel is held at fixed values. Results obtained from a device with a 5 nm channel on top of 90 nm $SiO_2$ gate dielectric are shown in Fig. 2b. When the gate voltage is varied from -30 V to 0 V, the channel switches from "on" state to "off" state, and a drop in drain current by a factor of ~ $10^5$ is observed. The measured drain current modulation is 4 orders of magnitude larger than that in graphene (due to its lack of bandgap), and approaches the value recently reported in MoS$_2$ devices[17]. Such a high drain current modulation makes black phosphorus thin film a promising material for applications in digital electronics[22]. Similar switching behaviour (with varying drain current modulation) is observed on all black phosphorous thin film transistors with thicknesses up to 50 nm. We note that the "on" state current of our devices has not yet reached saturation, due to the fact that the doping level is limited by the break-down electric field of the SiO$_2$ back-gate dielectric. It is therefore possible to achieve even higher drain current modulation by using high-k materials as gate dielectrics for higher doping. Meanwhile, a subthreshold swing (SS) of ~ 5 V/decade is observed, which is much larger than the SS in commercial Si-based devices (~ 70 mV/decade). We note that the SS in our devices varies from sample to sample (from ~ 3.7 V/decade to ~ 13.3 V/decade), and is on the same order of magnitude as reported in multilayer MoS$_2$ devices with a similar back-gate configuration[23,24]. The rather big SS is mainly attributed to the large thickness of the SiO$_2$ back-gate dielectric that we use, and multiple factors such as insulator layer thickness[25], Schottky barrier at subthreshold region[24], and sample-substrate interface state may be also at play.

The switching-off at the negative side of $V_g$ sweep is accompanied by a slight turn-on at positive gate voltages as shown in Fig. 2b. To further explore this ambipolar behaviour, we fabricate few-layer phosphorene devices with multiple electrical contacts (Fig. 2c inset), and perform Hall measurement using two opposing contacts (V$_2$ and V$_4$ for example) perpendicular to the drain-source current path to measure the transverse resistance, $R_{xy}$. The Hall coefficient $R_H$, defined as the slope of $R_{xy}$ as a function of external magnetic field $B$, reflects both the sign and the density of the charge carriers in the sample. As shown in Fig. 2c, a carrier sign inversion is clearly observed in the "on" states, with positive and negative gate voltage corresponding to hole and electron conduction respectively. This unambiguously shows that the ambipolar switching of the devices is caused by the Fermi level shifting from the valence band into the conduction band.

The nature of the electrical conduction is further probed by performing I-V measurement in a two-terminal configuration (Fig. 2a). As shown in fig. 2d, the source-drain current, $I_{ds}$, varies linearly with $V_{ds}$ in the "on" state of the hole side, indicating an ohmic contact in this region. Meanwhile $I_{ds}$ v.s. $V_{ds}$ is strongly non-linear on the electron side (Fig. 2d inset), typical for semiconducting channels with Schottky barriers at the contacts. The observed I-V characteristics can be readily explained by work function mismatch between the metal contacts and few-layer phosphorene: the high work function of the metal electrodes causes hole accumulation at the metal-semiconductor interface, which forms a low resistance ohmic contact for the p-doped sample; while for n-doped sample a depletion region is formed at the interface, leading to Schottky barriers and thus the non-linear conduction. This model also explains the observed disparity between the conduction at electron and the hole side in all our samples (Fig. 2b), and is widely accepted to describe the contact behaviour in $MoS_2$ devices[26].

For potential applications in digital and radio-frequency devices, the saturation of the drain current is crucial to reach the maximum possible operation speeds [22]. By carefully choosing the ratio between channel length and silicon dioxide layer thickness, a well-defined current saturation can be achieved in the high drain-source bias region (Fig. 3a). Meanwhile the electrical contacts remain ohmic in the linear region at low drain-source biases. Results shown in Fig. 3a are obtained in the "on" state of the hole side of the conduction in a 5 nm sample, with a 4.51-μm-long channel on 90 nm SiO$_2$ gate dielectric. Such a well-developed saturation behaviour, which is absent in graphene based FET devices[22], is crucial for reaching high power gains. Coupled with the fact that our channel thickness is on the order of nm, and thus robust against short-channel effects when the channel length is shrunk to nm scale, our results suggest the high potential of black phosphorus in high speed field-effect devices application. We note that the "on" state conductance of our device is relatively low and the threshold source-drain bias is relatively high compared to typical Si-based devices. Both of these two factors are attributed to the long channel length in our current device. Better device performance, i.e. larger saturation current and lower threshold bias, is expected if the channel length and the gate oxide thickness are reduced. Further investigations are needed to test the limit of the device performances of black phosphorus FETs.

We now turn to the characterization of field-effect mobility in few-layer phosphorene devices. Conductance, $G$, was measured as a function of $V_g$, and we extract the field-effect mobility, $\mu_{FE}$, in the linear region of the transfer characteristics[27]:

$$\mu_{FE} = \frac{L}{W} \frac{1}{C_g} \frac{dG}{d(V_g - V_{th})} \tag{1}$$

where $L$ and $W$ are the length and width of the channel respectively, $C_g$ is the capacitance per unit area, and $V_{th}$ the threshold gate voltage. Hole mobility as high as 984 $cm^2V^{-1}s^{-1}$ is obtained on a 10 nm sample as shown in Fig. 3b, and is found to be strongly thickness-dependent. Transfer characteristics of two other typical samples of different thicknesses (8 nm and 5 nm, with the 5 nm sample the same one measured in Fig. 3a) are shown in Fig. 3b. The conductance is measured in a four-terminal configuration to avoid complications from electrical contacts[25]. Two-terminal conductance measurement setup is also used on some of our devices. It is found to over-estimate the hole mobility, but still yields values on the same order of magnitude (Fig. 3b inset). Such mobility values, though still much lower than that in graphene[28–30], compare favorably with $MoS_2$ samples[17,18,24], and are already much higher than values found in typical silicon-based devices commercially available (~ 500 $cm^2/Vs$ )[22].

The thickness dependence of the two key metrics of material performance – drain current modulation and mobility – is further explored to elucidate the transport mechanism of few-layer phosphorene FETs. The experimental results are summarized in Fig. 3b inset. The drain current modulation decreases monotonically as the sample thickness is increased, while the mobility peaks at ~10 nm and comes down slightly beyond that. Similar thickness-dependence of carrier mobility has been reported in other 2D FETs such as few-layer graphene and $MoS_2$, where models taking into account the screening of gate electric field were invoked to account for the observed behaviour[32,33]. Simply speaking, the gate electric field induces free carriers only in the bottom layers due to charge screening. So the top layers still give finite conduction in the "off" state, reducing the drain current modulation. Meanwhile field-effect mobility is also dominated by the contribution from layers at the bottom. Thinner samples are more susceptible to the charge impurities at the interface (thus their lower mobilities) which are otherwise screened by the induced charge in thicker samples. This explains the sharp increase of the field-effect mobility below ~ 10 nm. As the samples get thicker, however, another factor has to be taken into account: since the current is injected from electrical contacts on the top surface, the finite inter-layer resistance forces the current to flow in the top layers which are not gated by the back-gate. This effect depresses the field-effect mobility for samples thicker than ~ 10 nm. Based on above arguments, we model the electrical conduction in our samples using a self-consistently obtained carrier distribution (for more details see Supplementary Information) and our calculation fits well with the experimental data as shown in Fig. 3b inset. The model also suggests a way to achieve higher mobility without sacrificing the drain current modulation: using a top-gate device structure with a layer of high-k dielectric material as the gate dielectric, one could effectively screen the charge impurities, but leave the drain current modulation intact. In addition, since the top layers where the current flows are now gated by the top-gate, the mobility is no longer affected by the inter-layer resistance. Such a method has been proven to work in $MoS_2$ FETs[17,18].

Finally we examine the temperature dependence of the carrier mobility to uncover various factors that limit the mobility in our FETs. Here two types of carrier mobility are measured on the same device for comparison: one is the $\mu_{FE}$ extracted from the linear part of the gate-dependent conductance (Fig. 4a) according to Eq. (1); and the other is Hall mobility $\mu_H$ obtained by

$$\mu_H = \frac{L}{W}\frac{G}{ne} \qquad (2)$$

where $e$ is the charge of an electron and $n$ is the 2D charge density determined from gate capacitance $n = C_g(V_g - V_{th})$, which equals to the density extracted from Hall measurement $n = 1/eR_H$ if the sample geometry permits an accurate determination of Hall coefficient $R_H$ (see Supplementary Information for details). The two mobilities in an 8 nm sample as functions of temperature are shown in Fig. 4b. They fall in the vicinity of each other, and show similar trend as the temperature is varied: both decrease at temperature higher than ~ 100 K, and saturate (or decrease slightly for low carrier densities) at lower temperatures. The behaviour of the mobility as the temperature is lowered to 2 K is consistent with scattering from charged impurities[31]. We note that in this temperature range the Hall mobility increases as the gate induced carrier density gets larger (Fig. 4d). The reduced scattering in the sample points to the diminished disorder potential as a result of screening by free charge carriers. This further corroborates our model that the charged impurity at the sample/substrate interface is a limiting factor for the carrier mobility. On the other hand the drop in mobility from ~100 K up to 300 K can be attributed to electron-phonon scattering that dominates at high temperatures[31], and the temperature dependence roughly follows a power law $\mu \propto T^{-\gamma}$ as seen in Fig. 4b. The exponent $\gamma$ depends on the electron-phonon coupling in the sample, and is found close to ~ 0.5 in our 8 nm device (as a guide to the eye, $\mu \sim T^{-0.5}$ is plotted in Fig. 4b as a dashed line). This $\gamma$ value for few-layer phosphorene is notably smaller than the value in other 2D materials[32] and bulk black phosphorus[11], but agrees with that in monolayer $MoS_2$ covered by a layer of high-κ dielectric[18]. The exact mechanism of the suppression of phonon scattering in few-layer phosphorene is not clear at this moment and deserves further study.

In conclusion, we have succeeded in fabricating p-type FETs based on few-layer phosphorene. Our samples exhibit ambipolar behavior with drain current modulation up to ~$10^5$, and a field effect mobility value up to ~ 1000 $cm^2V^{-1}s^{-1}$ at room temperature. The carrier mobility is limited by charge impurity scattering at low temperatures and electron-phonon scattering at high temperatures. The on current is low and subthreshold swing is high, but optimization of the gate dielectric should improve these characteristics. The ability to make transistors combined with the fact that few-layer phosphorene has a direct band gap in the infrared regime, make black phosphorus a candidate for future nano-electronic and opto-electronic applications.

**Methods**

Sample growth

Black phosphorus was synthesized under a constant pressure of 10 kbar by heating red phosphorus to 1000 ℃ and slowly cooling to 600 ℃ at a cooling rate of 100 ℃ per hour. Red phosphorus was purchased from Aladdin Industrial Corporation with 99.999% metals basis. High pressure environment was provided by a cubic-anvil-type apparatus (Riken CAP-07). X-ray diffraction (XRD) was performed on a Smartlab-9 diffractometer (Rikagu) using Cu Kα radiation (Fig. S1, Supplementary Information).

Measurements

The ARPES measurements were performed at BaDElPh beamline at the Elettra synchrotron radiation facility with an MBS A-1 electron analyzer. The overall energy resolution was set to 20 meV or better and the typical angular resolution is 0.5 Deg. During the measurement the temperature was kept at 60 K to avoid the onset of charging. Data shown in Fig. 1b were taken with s-polarized 20 eV photons and no obvious polarization dependent but noticeable intensity variation were observed for observed bands.

Transport measurements are mainly performed in an Oxford Instruments Optistat AC-V12 cryostat with samples in vacuum (~ $10^{-5}$ mBar). Part of measurements is done in the Oxford Instruments Integra$^{TM}$ AC cryostat and Quantum Design PPMS when magnetic field is needed. Data are collected in a DC setup using DL 1211 current preamplifier with a voltage source, or Keithley 6220 current source combined with Keithley 2182 nanovoltmeter. Some of the Hall measurements are done using an SRS 830 lock-in amplifier.

Band structure calculation

Our *ab initio* band structure calculations, based on density functional theory, were performed using the projector augmented wave method[33,34], as implemented in the Vienna ab initio Simulation Package (VASP) code[35]. The crystal structure data of black phosphorus were taken from References[2,7]. For the exchange-correlation energy, we used the screened hybrid density functional of the Heyd-Scuseria-Ernzerhof type (HSE06)[36]. Details of the calculation are described in the Supplementary Information.

**Acknowledgements**

We thank Ruibao Tao, Feng Wang, Yanqing Wu, Liguo Ma, Mengqiao Sui, Guorui Chen and Fangyuan Yang for helpful discussions, Faxian Xiu, Yanwen Liu and Cheng Zhang for assistance with measurements in PPMS, and Dr. Luca Petaccia for the experimental support at BaDElPh beamline. Part of the sample fabrication was performed at Fudan Nano-fabrication Lab. L.L., Y.Y., Q.G, D.F. and Y.Z. acknowledge financial support of the National Basic Research Program of China (973 Program) under the grant Nos. 2011CB921802, 2012CB921400 and 2013CB921902, and NSF of China under the grant No. 11034001. G.J.Y and X.H.C. acknowledge support from the 'Strategic Priority Research Program' of the Chinese Academy of Sciences under the grant No. XDB04040100 and the National Basic Research Program of China (973 Program) under the grant No. 2012CB922002. X.O. and H.W. are supported by Pu Jiang Program of Shanghai under the grant No. 12PJ1401000.

**Author contributions**

X.H.C. and Y.Z. conceived the project. G.J.Y. and X.H.C. grew bulk black phosphorus crystal. L.L. fabricated black phosphorus thin film devices and performed electric measurements, and L.L., Y.Y. and Y.Z. analyzed the data. Q.G. and D.F. did ARPES measurement on bulk black phosphorus crystal. X.O. and H.W. carried out *ab initio* band structure calculations. L.L. and Y.Z. wrote the paper and all authors commented on it.

**Additional information**

Supplementary information is available in the online version of the paper. Reprints and permissions information is available online at http://npg.nature.com/reprintsandpermissions. Correspondence and requests for materials should be addressed to X.H.C. or Y.Z.

**Figure captions**

**Figure 1. Crystal and electronic structure of bulk black phosphorus. a,** Atomic structure of black phosphorus. **b,** Band structure of bulk black phosphorus mapped out by ARPES measurement. A band gap is clearly observed. Superimposed on top are calculated bands of bulk crystal. Solid and dash lines are empty and filled bands respectively. The directions of the ARPES mapping are along **U** (**L**-**Z**) and **T'**, as indicated in the first Brillion zone shown in the inset.

**Figure 2. Few-layer phosphorene FET and its device characteristics. a,** Schematic of the device structure of a few-layer phosphorene FET. The device profile shown here is the three-dimensional rendering of the AFM data. Electrodes and few-layer phosphorene crystal are false-colored to match how they appear under microscope. Cross-section of the device along the white dash line is shown in the bottom graph. **b,** Source-drain current (in logarithmic scale) as a function of gate voltage obtained from a 5-nm-thick device on Si substrate with 90 nm $SiO_2$ at room temperature. Drain current modulation up to ~ $10^5$ is observed for both drain-source biases on the hole

side of the gate doping. A slight turn-on at the electron side is also observed. **c,** Hall coefficient (blue curve) and conductance (red curve) as a function of gate voltage collected from a 8-nm-thick sample on Si substrate with 285 nm $SiO_2$. Carrier type inversion, signified by the sign change of the Hall coefficient, are observed when the polarity of the gate is reversed. **d,** I-V characteristics for the same device shown in Fig. 2b. Linear behaviour indicates ohmic contact on the hole side of the gate doping. Meanwhile the non-linear behaviour on the hole side of the gate doping (inset) indicates the formation of Schottky barriers at the contacts.

**Figure 3. Current saturation and mobitliy of few-layer phosphorene FET. a,** Drain-source current $I_{ds}$ as a function of bias $V_{ds}$ at different gate voltages collected from a 5-nm-thick device on Si substrate with 90 nm $SiO_2$. A saturation region is observed for all gate voltages applied. **b,** Sheet conductivity measured as a function of gate voltage for devices with different thicknesses: 10 nm (black solid line), 8 nm (red solid line) and 5 nm (green solid line). The 5 nm device is the same one measured in **a**. Field-effect mobility is extracted from the line fit of the linear region of the conductivity (dash lines). All gate voltages are normalized to 90 nm gate oxide for easy comparison. The inset summaries the drain current modulation (filled blue triangles) and carrier mobility (open circles) of black phosphorus FET with varying thicknesses. Mobilities measured in 4-terminal and 2-terminal configuration are denoted by black and red circles respectively. Dashed lines are the models described in the main text.

**Figure 4. Temperature-dependent behaviour of few-layer phosphorene FET. a**, Four-terminal conductance as a function of gate voltage measured in a 8-nm-thick sample at different temperatures. **b,** Field-effect mobility (open circles) and Hall mobility (filled squares) as a function of temperature in logarithm scale. Data are extracted from the same sample in **a.** A power law dependence $\mu \sim T^{-0.5}$ is plotted in the high temperature region as a guide to the eye.

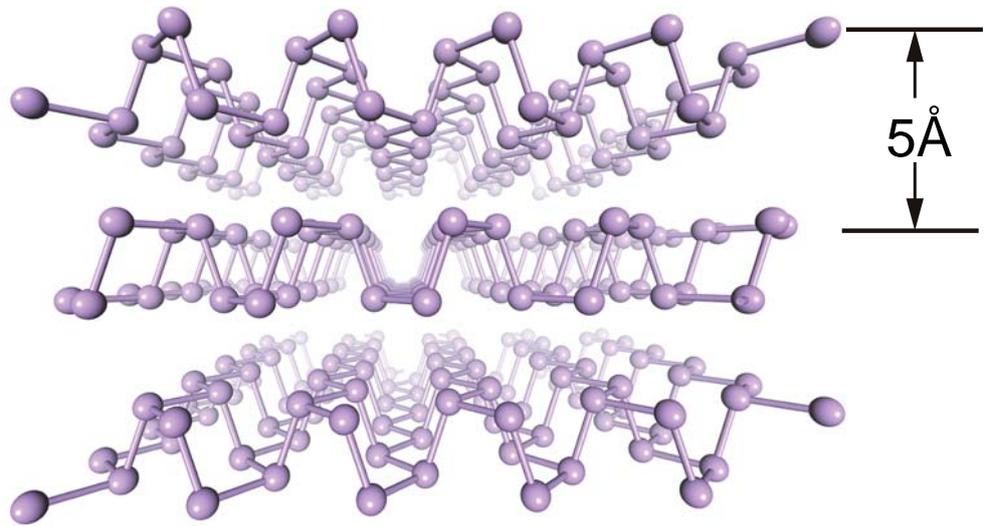

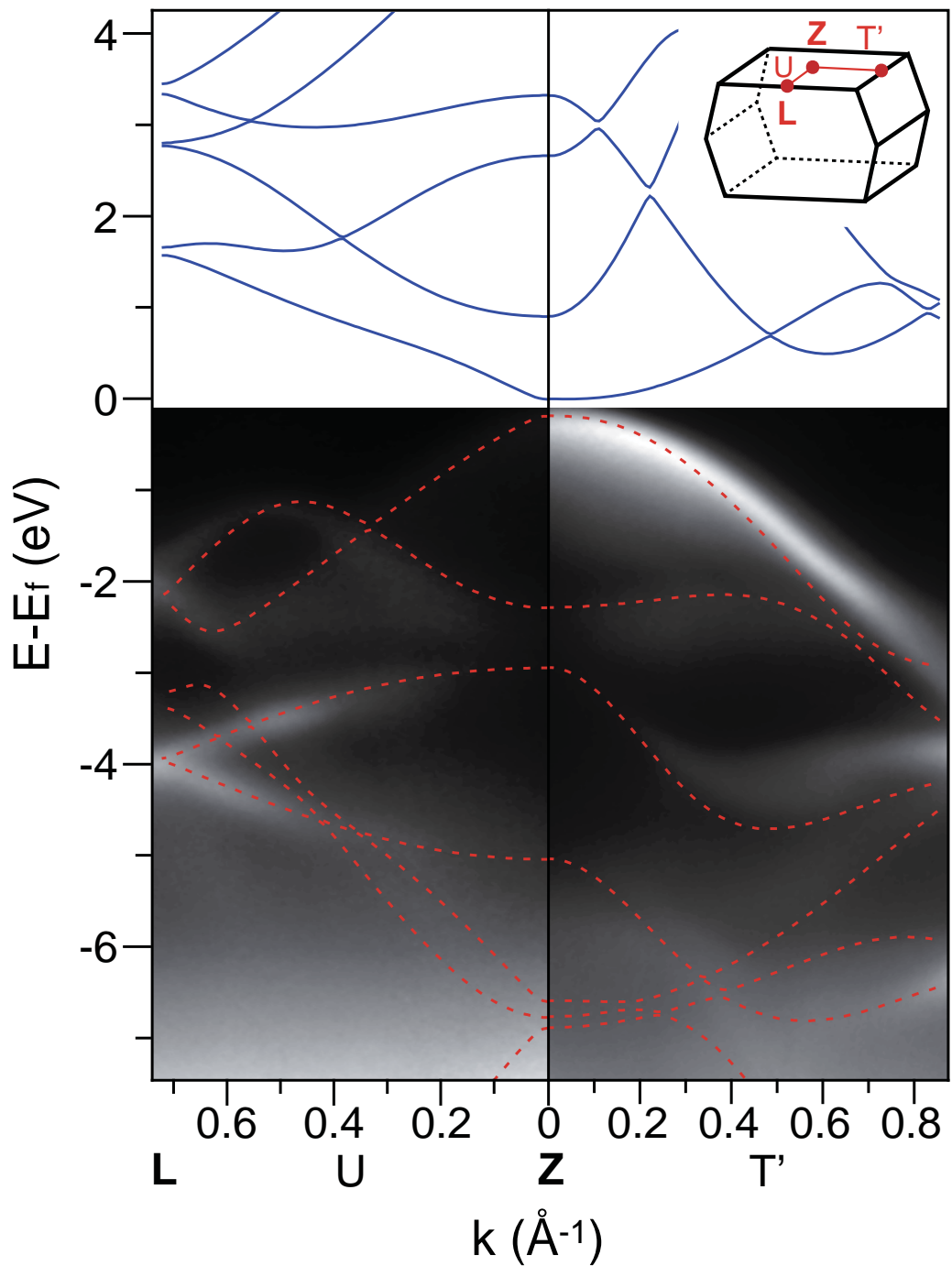

L. Li et al.　　　　　　　　　　　　　　　　　　　　　　　　　　　　Figure 1

a

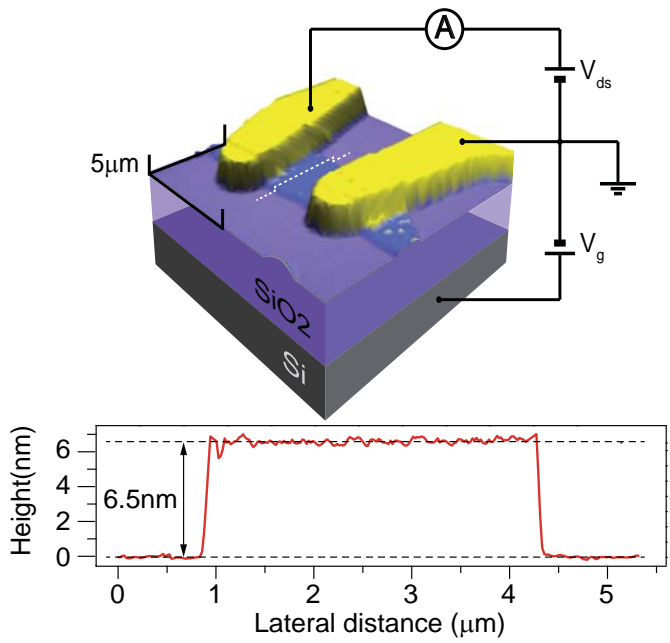

b

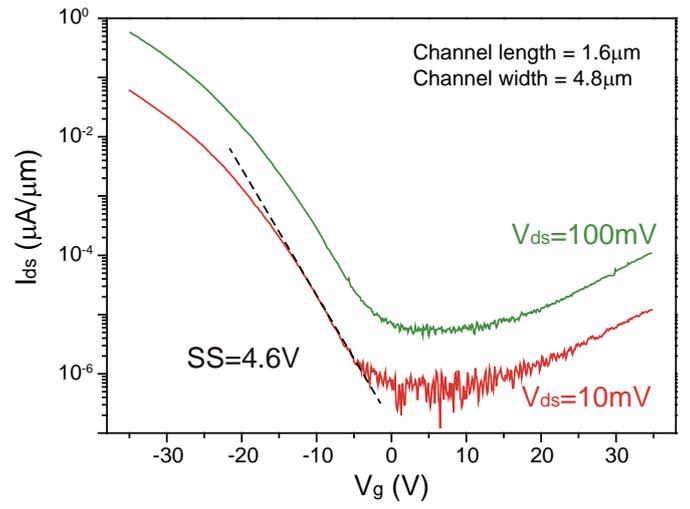

c

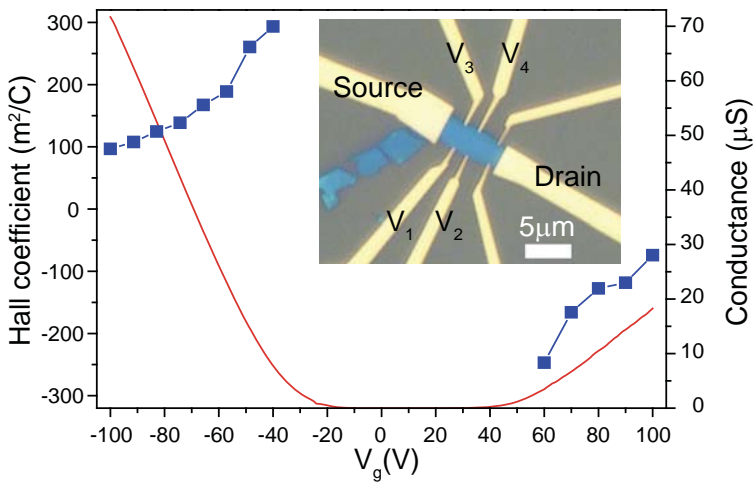

d

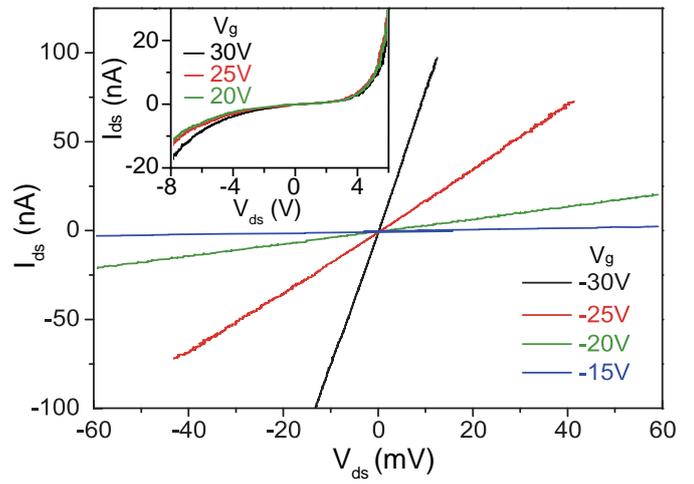

L. Li et al.  Figure 2

a

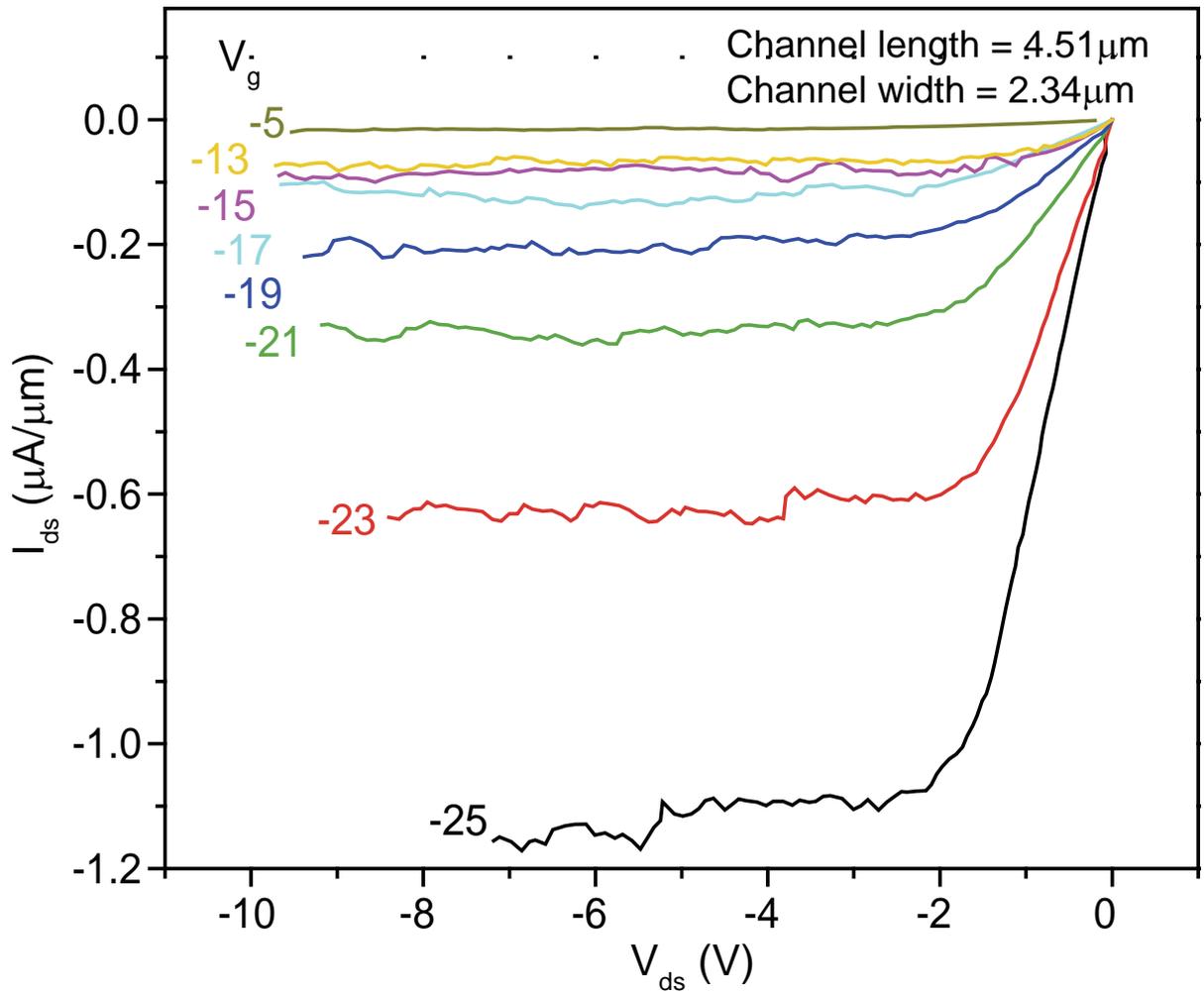

b

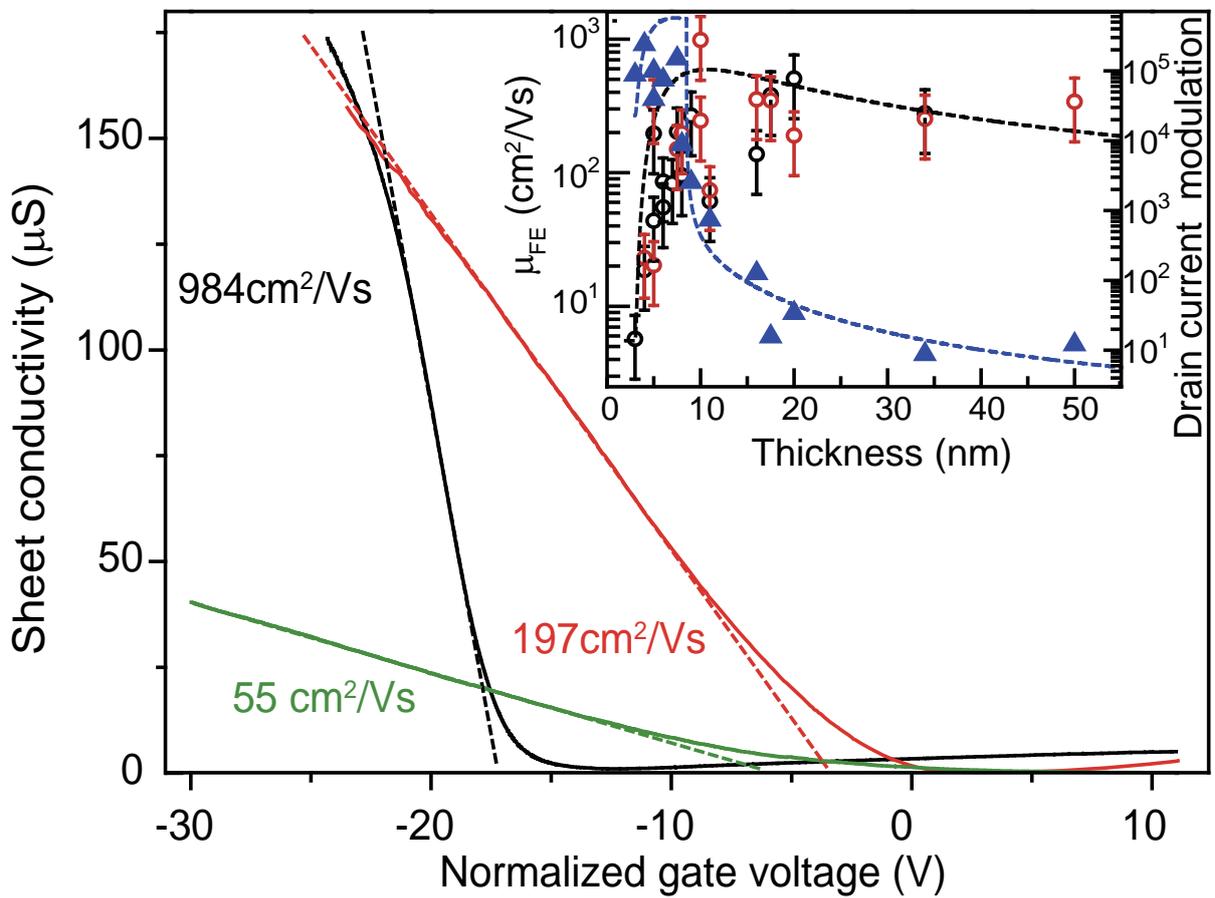

L. Li et al.

Figure 3

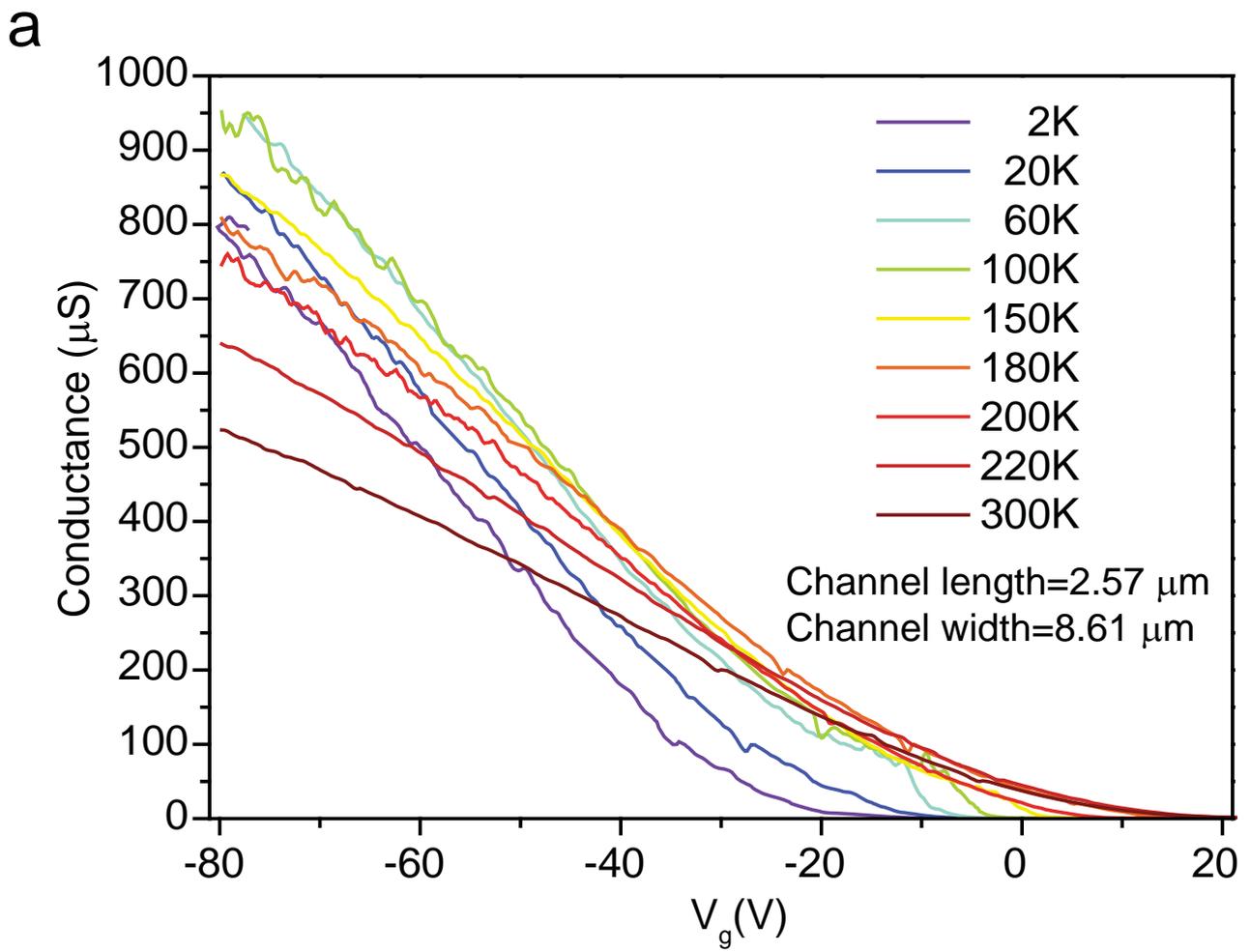

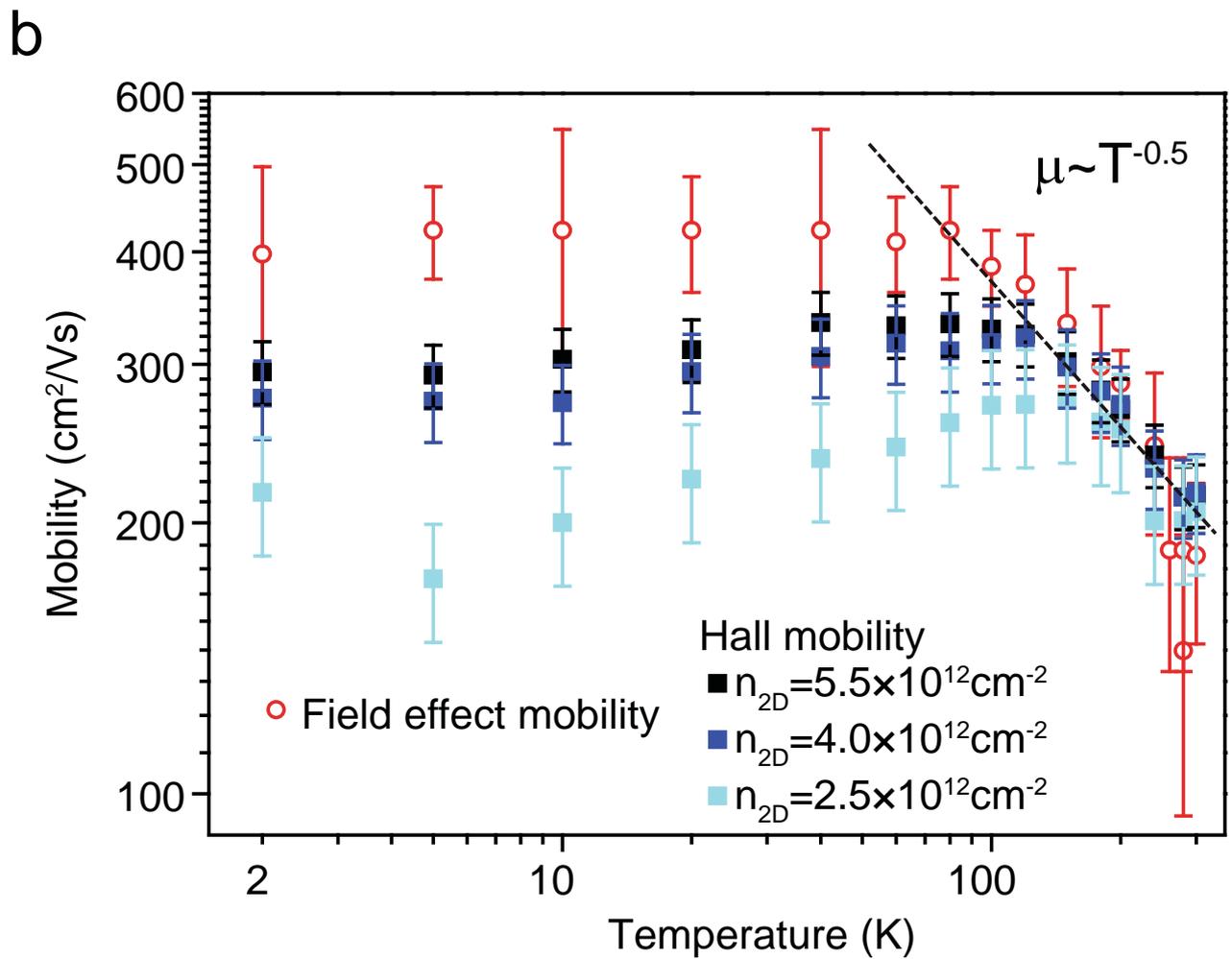